\documentclass[twocolumn,aps,prl,a4paper,preprintnumbers,a4paper,superscriptaddress,showpacs,tightenlines,amsmath,amssymb]{revtex4}
\usepackage{graphicx} 
\usepackage{dcolumn}  

\graphicspath{{ps}}

\begin{document}


\title{ \quad\\[1.0cm] Improved measurement of time-dependent $CP$ violation 
in $B^0\rightarrow J/\psi\, \pi^0$ decays}

\affiliation{Budker Institute of Nuclear Physics, Novosibirsk}
\affiliation{Chiba University, Chiba}
\affiliation{University of Cincinnati, Cincinnati, Ohio 45221}
\affiliation{The Graduate University for Advanced Studies, Hayama}
\affiliation{Gyeongsang National University, Chinju}
\affiliation{Hanyang University, Seoul}
\affiliation{University of Hawaii, Honolulu, Hawaii 96822}
\affiliation{High Energy Accelerator Research Organization (KEK), Tsukuba}
\affiliation{Institute of High Energy Physics, Chinese Academy of Sciences, Beijing}
\affiliation{Institute of High Energy Physics, Vienna}
\affiliation{Institute of High Energy Physics, Protvino}
\affiliation{Institute for Theoretical and Experimental Physics, Moscow}
\affiliation{J. Stefan Institute, Ljubljana}
\affiliation{Kanagawa University, Yokohama}
\affiliation{Korea University, Seoul}
\affiliation{Kyungpook National University, Taegu}
\affiliation{Ecole Polyt\'ecnique F\'ed\'erale Lausanne, EPFL, Lausanne}
\affiliation{University of Ljubljana, Ljubljana}
\affiliation{University of Maribor, Maribor}
\affiliation{University of Melbourne, School of Physics, Victoria 3010}
\affiliation{Nagoya University, Nagoya}
\affiliation{Nara Women's University, Nara}
\affiliation{National Central University, Chung-li}
\affiliation{National United University, Miao Li}
\affiliation{Department of Physics, National Taiwan University, Taipei}
\affiliation{H. Niewodniczanski Institute of Nuclear Physics, Krakow}
\affiliation{Nippon Dental University, Niigata}
\affiliation{Niigata University, Niigata}
\affiliation{University of Nova Gorica, Nova Gorica}
\affiliation{Osaka City University, Osaka}
\affiliation{Osaka University, Osaka}
\affiliation{Panjab University, Chandigarh}
\affiliation{Saga University, Saga}
\affiliation{University of Science and Technology of China, Hefei}
\affiliation{Seoul National University, Seoul}
\affiliation{Sungkyunkwan University, Suwon}
\affiliation{University of Sydney, Sydney, New South Wales}
\affiliation{Tata Institute of Fundamental Research, Mumbai}
\affiliation{Toho University, Funabashi}
\affiliation{Tohoku Gakuin University, Tagajo}
\affiliation{Tohoku University, Sendai}
\affiliation{Department of Physics, University of Tokyo, Tokyo}
\affiliation{Tokyo Institute of Technology, Tokyo}
\affiliation{Tokyo Metropolitan University, Tokyo}
\affiliation{Tokyo University of Agriculture and Technology, Tokyo}
\affiliation{Virginia Polytechnic Institute and State University, Blacksburg, Virginia 24061}
\affiliation{Yonsei University, Seoul}
 \author{S.~E.~Lee}\affiliation{Seoul National University, Seoul} 
 \author{K.~Miyabayashi}\affiliation{Nara Women's University, Nara} 
  \author{H.~Aihara}\affiliation{Department of Physics, University of Tokyo, Tokyo} 
  \author{T.~Aushev}\affiliation{Ecole Polyt\'ecnique F\'ed\'erale Lausanne, EPFL, Lausanne}\affiliation{Institute for Theoretical and Experimental Physics, Moscow} 
  \author{T.~Aziz}\affiliation{Tata Institute of Fundamental Research, Mumbai} 
  \author{A.~M.~Bakich}\affiliation{University of Sydney, Sydney, New South Wales} 
  \author{V.~Balagura}\affiliation{Institute for Theoretical and Experimental Physics, Moscow} 
  \author{E.~Barberio}\affiliation{University of Melbourne, School of Physics, Victoria 3010} 
 \author{A.~Bay}\affiliation{Ecole Polyt\'ecnique F\'ed\'erale Lausanne, EPFL, Lausanne} 
  \author{K.~Belous}\affiliation{Institute of High Energy Physics, Protvino} 
% ...
\collaboration{The Belle Collaboration}
\noaffiliation

\begin{abstract}
We report improved measurements of time-dependent 
\emph{CP} violation parameters for 
$B^0(\overline{B}^0) \to J/\psi\, \pi^0$ decay.
This analysis is based on 535 million $B\overline{B}$ pairs 
accumulated at the $\Upsilon(4S)$ resonance with the Belle detector at 
the KEKB asymmetric-energy $e^+e^-$ collider.
From the distribution of proper time intervals between the two $B$ 
decays, we obtain the following $CP$ violation parameters
\begin{center}
$\mathcal{S}_{J/\psi\, \pi^0} 
= -0.65\pm0.21 (\rm{stat})\pm0.05 (\rm{syst})$ and\\
$\mathcal{A}_{J/\psi\, \pi^0} 
= +0.08\pm0.16 (\rm{stat})\pm0.05 (\rm{syst})$,
\end{center}
which are consistent with Standard Model expectations.
\end{abstract}


\maketitle

\tighten

{\renewcommand{\thefootnote}{\fnsymbol{footnote}}}
\setcounter{footnote}{0}

The Kobayashi-Maskawa (KM) quark-mixing matrix~\cite{KM}
has an irreducible complex phase that gives rise to 
$CP$-violating asymmetries in the time-dependent rates 
of $B^0$ and $\overline{B}{}^0$ decays 
into a common $CP$ eigenstate such as $J/\psi\, \pi^0$~\cite{carter}.
In the decay chain
$\Upsilon$(4S)$\to B^0\overline{B}{}^0\to$ $(J/\psi\, \pi^0) f_{\rm tag}$, 
where one of the $B$ mesons decays at time $t_{CP}$ to the final state 
$J/\psi\, \pi^0$
and the other decays at time $t_{\rm tag}$ to 
a final state 
$f_{\rm tag}$ that
distinguishes between $B^0$ and $\overline{B}{}^0$,
the decay rate has a time dependence given by~\cite{cpviolation}
\begin{eqnarray}
\label{signal_cp}
{\cal P}(\Delta{t})= \frac{ e^{-|\Delta{t}|/{\tau_{B^0}}} }{4\tau_{B^0}}
\biggl\{1 & + & q \cdot
 \Bigl[ {\cal S}_{J/\psi\, \pi^0} \sin(\Delta m_d \Delta{t})  \nonumber \\
 & +  & {\cal A}_{J/\psi\, \pi^0} \cos(\Delta m_d \Delta{t})
\Bigr] \biggr\},
\end{eqnarray}
where $\tau_{B^0}$ is 
the neutral $B$ lifetime,
$\Delta m_d$ is the mass difference between 
the two neutral $B$ mass eigenstates, 
$\Delta t = t_{CP}-t_{\rm tag}$, and
the $b$-flavor charge $q = +1~(-1)$ when the tagging $B$ meson is 
a $B^0(\overline{B}{}^0)$.
The $CP$ violation parameters $\mathcal{S}_{J/\psi\, \pi^0}$ and 
$\mathcal{A}_{J/\psi\, \pi^0}$ are given by
\begin{equation}
\mathcal{S}_{J/\psi\, \pi^0}\equiv \frac{2\Im(\lambda)}{|\lambda|^2+1},		
\hspace{0.2in}\mathcal{A}_{J/\psi\, \pi^0}\equiv \frac{|\lambda|^2-1}{|\lambda|^2+1
\label{para_lambda}
}
\end{equation}
where $\lambda$ is a complex parameter that depends on both the 
$B^0\overline{B}{}^0$ mixing and the amplitudes 
for $B^0$ and $\overline{B}{}^0$ decay to $J/\psi\, \pi^0$.
In the Standard Model~(SM), $|\lambda|$ is, to a good
approximation, equal to
the absolute value of the ratio of 
the $\overline{B}{}^0\to J/\psi\, \pi^0$ to $B^0\to J/\psi\, \pi^0$
decay amplitudes.
At the quark level, the $B^0\to J/\psi\, \pi^0$ decay proceeds
via a $b\to c\overline{c}d$ transition.
In this decay,
the tree amplitude is
CKM-suppressed.
Since the tree amplitude has the same weak phase 
as the $b\to c\overline{c}s$ transition,
$\mathcal{S}_{J/\psi\, \pi^0} = -\sin2\phi_{1}$
and $\mathcal{A}_{J/\psi\, \pi^0} = 0$ are expected if other
contributions to the decay amplitude can be neglected~\cite{grossman}.
If, however, the penguin or other contributions are substantial, 
the $CP$ violation parameters for this mode
may deviate from these values.
Employing SU(3) symmetry as well as plausible dynamical
assumptions, the results obtained for
$B \to J/\psi\, \pi^0$ decay can be used to estimate the penguin pollution
in $B^0 \to J/\psi\, K^0_S$ decay for a very precise
determination of $\sin 2 \phi_1$~\cite{ciuchini}.

The most recent study of $B^0\to J/\psi\, \pi^0$ decays was 
reported by BaBar~\cite{babar}
using a sample of 232 million $B\overline{B}$ pairs,
while the previous Belle analysis~\cite{kataokaPRL}
was based on a data sample corresponding to 152 million $B\overline{B}$ 
pairs. This measurement of time-dependent $CP$ violation
in $B^0\to J/\psi\, \pi^0$ decays is based on a larger data 
sample that contains 535 million $B\overline{B}$ pairs, 
collected  with the Belle detector at the KEKB asymmetric-energy
$e^+e^-$ (3.5 on 8~GeV) collider~\cite{KEKB}
operating at the $\Upsilon(4S)$ resonance.
The $\Upsilon(4S)$ is produced with a Lorentz boost factor
of $\beta\gamma = 0.425$ along the $z$-axis, 
which is anti-parallel to the positron beam direction.
Since the $B\overline{B}$ pairs are produced nearly at rest
in the $\Upsilon(4S)$ center-of-mass system (cms),
$\Delta t$ is determined from $\Delta z$, the distance 
between the two $B$ meson decay vertices along 
the $z$-direction: $\Delta t \cong \Delta z/c\beta\gamma$, 
where $c$ is the speed of light.

The Belle detector is a large-solid-angle magnetic
spectrometer that
consists of a silicon vertex detector (SVD),
a 50-layer central drift chamber (CDC), an array of
aerogel threshold Cherenkov counters (ACC), 
a barrel-like arrangement of time-of-flight
scintillation counters (TOF), and an electromagnetic calorimeter
comprised of CsI(Tl) crystals (ECL) located inside 
a superconducting solenoid coil that provides a 1.5~T
magnetic field.  An iron flux-return located outside
the coil is instrumented to detect $K_L^0$ mesons and to identify
muons (KLM).  The detector
is described in detail elsewhere~\cite{Belle}.
Two inner detector configurations were used. 
A 2.0 cm radius beam pipe
and a 3-layer silicon vertex detector were used for the first sample
of 152 million $B\overline{B}$ pairs, 
while a 1.5 cm radius beampipe, a 4-layer
silicon detector and a small-cell inner drift chamber were used to record  
the remaining 383 million $B\overline{B}$ pairs~\cite{svd2}.

We reconstruct $J/\psi$ mesons in the $\ell^+ \ell^-$ decay channel
($\ell = e$ or $\mu$) and include up to two bremsstrahlung photons
that are within 50 mrad of each of the $e^+$ and $e^-$ tracks (denoted
as $e^+ e^- (\gamma)$). The invariant mass is required to be 
within 
$-0.15$ GeV/$c^2 < M_{ee(\gamma)} - m_{J/\psi} < +0.036$ GeV/$c^2$ 
and
$-0.06$ GeV/$c^2 < M_{\mu \mu} - m_{J/\psi} < +0.036$ GeV/$c^2$, 
where $m_{J/\psi}$ denotes the $J/\psi$ nominal mass~\cite{pdg2006},
and $M_{ee(\gamma)}$ and $M_{\mu \mu}$ are the reconstructed invariant
masses from $e^+ e^- (\gamma)$ and $\mu^+ \mu^-$, respectively.

Photon candidates are selected from clusters 
of up to $5\times5$ crystals in the ECL.
Each candidate is required to have no associated charged track 
and a cluster shape that is consistent with an electromagnetic shower.
To select $\pi^0\to \gamma\gamma$ decay candidates,
the energy of a photon is required to be greater than 
50 MeV in the ECL barrel and 100 MeV in the end-cap region. 
A pair of photons with an invariant mass in the range
118 MeV/$c^2$ $< M_{\gamma\gamma} <$ 150 MeV/$c^2$ is 
considered as a $\pi^0$ candidate.

We combine the $J/\psi$ and $\pi^0$ to form a neutral $B$ meson. Signal
candidates are identified by two kinematic variables defined
in the $\Upsilon(4S)$ rest frame (cms): the beam-energy constrained mass 
$M_{\rm bc}\equiv \sqrt{E^{2}_{\rm beam}-(\sum \overrightarrow{p_i})^2}$ and 
the energy difference $\Delta E \equiv \sum E_i - E_{\rm beam}$, 
where $E_{\rm beam}=\sqrt{s}/2$ is the cms beam energy, 
and $\overrightarrow{p_i}$ and $E_i$ are the cms three momenta
and energies of the candidate $B$ meson decay products, respectively.
In order to improve the $\Delta E$ resolution,
vertex- and mass-constrained fits are applied to
$J/\psi\to \ell^+\,\ell^-$
decays and a mass constrained fit is used for
$\pi^0\to\gamma \gamma$ decays.
The $B$ meson signal region is defined as 
5.27 GeV/$c^2 < M_{\rm bc} <$ 5.29 GeV/$c^2$ and
$-0.1$ GeV $< \Delta E <$ 0.05 GeV.
The lower bound in $\Delta E$ is chosen to accommodate 
the negative $\Delta E$ tail of the signal due to shower leakage 
associated with the $\pi^0$, and to avoid background from
$B^0 \to J/\psi\, K^0_S$ $(K^0_S \to \pi^0\, \pi^0)$ decays.
To suppress the two-jet-like $e^+ e^- \to q \overline{q}$ 
$(q = u, d, s, c)$ continuum background, we require that the event
shape variable, $R_2$, which is the ratio of the second to zeroth Fox-Wolfram 
moment, satisfy 
$R_2 < 0.4$~\cite{foxwolf}.

We identify the flavor of the accompanying $B$ meson 
from inclusive properties of particles
that are not associated with the reconstructed $B^0\to J/\psi\, \pi^0$.
The algorithm for flavor tagging is described 
in detail elsewhere~\cite{tagging}.
We use two parameters, $q$ defined in Eq.(\ref{signal_cp}) and $r$,
to represent the tagging information.
The parameter $r$ is an event-by-event Monte Carlo (MC) determined 
flavor-tagging quality factor that ranges from $r=0$ for no flavor 
discrimination to $r=1$ for unambiguous flavor assignment. 
It is used only for sorting data into six intervals. 
The wrong tag fractions for the six $r$ intervals, $w_l$ $(l=1,6)$, 
and the difference in $\omega$ between $B^0$ and $\overline{B}{}^0$ decays, 
$\Delta w_l$, are determined from data~\cite{tagging}.
The vertex position for the $J/\psi\, \pi^0$ decay is reconstructed using
leptons from the $J/\psi$ decay. 
The vertex position of $f_{\rm tag}$ is obtained using tracks that
are not assigned to the $J/\psi\, \pi^0$ candidate and an interaction point constraint.
After all selection criteria are applied,
we obtain 864 events in the $\Delta E\!-\!M_{\rm bc}$ fit region
defined as $5.2~\mbox{GeV/}c^2 < M_{\rm bc} < 5.3~\mbox{GeV/}c^2$
and $-0.2~\mbox{GeV} < \Delta E < 0.2~\mbox{GeV}$, of
which 290 are in the signal box.

We perform an unbinned maximum likelihood fit to the 
$\Delta E\!\!-\!\!M_{\rm bc}$ distribution in order to distinguish
signal and backgrounds.
\begin{figure}[htb]
\includegraphics[width=7.5cm]{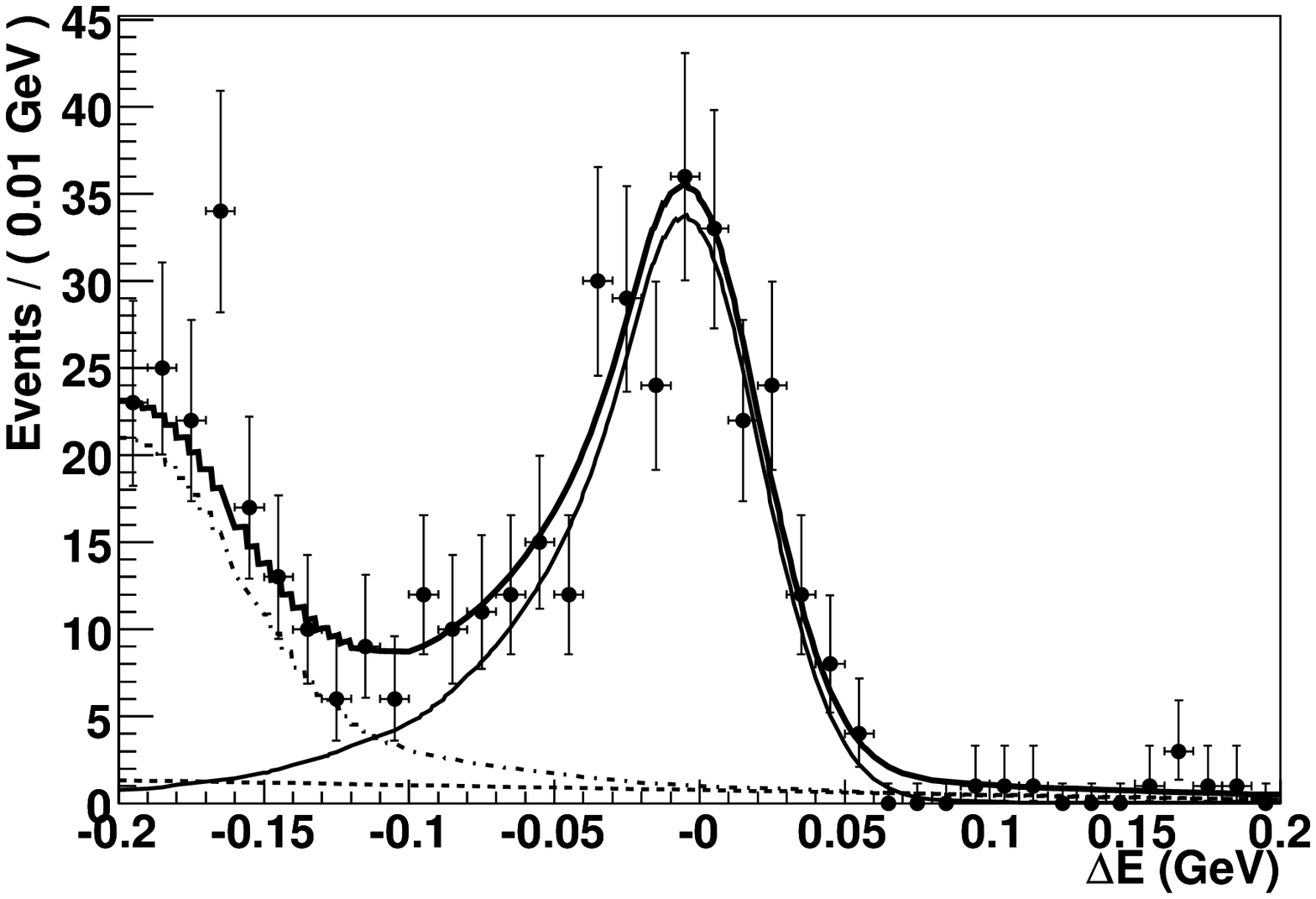}
\hfill
\includegraphics[width=7.5cm]{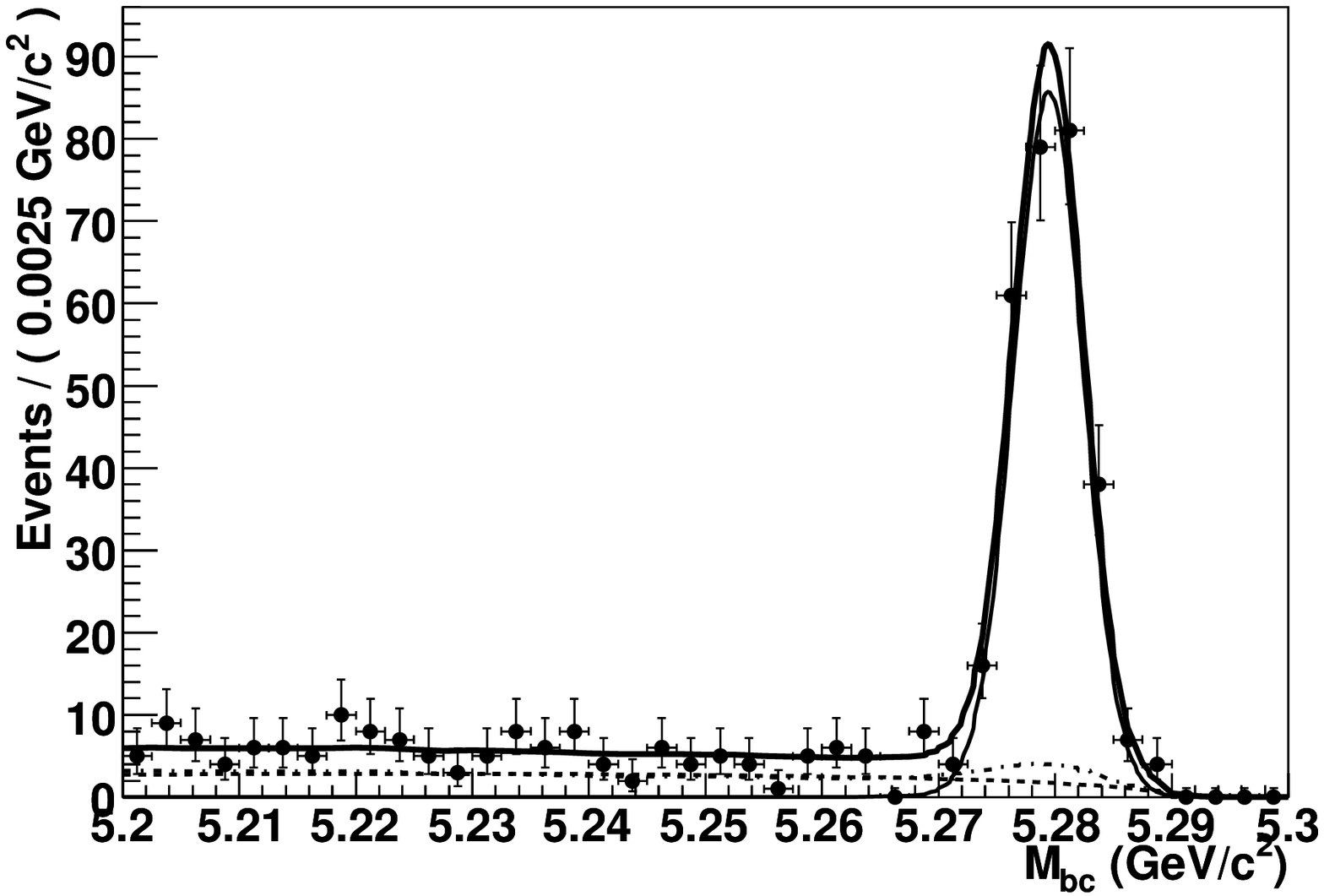}
\caption{$\Delta E$ distribution for events in the
$M_{\rm bc}$ signal region (top) and $M_{\rm bc}$ distribution
in the $\Delta E$ signal region (bottom).
The superimposed curves show the signal (solid line), $B\to J/\psi\, X$
(dot-dashed line), combinatorial background (dashed line) and
the sum of all the contributions (thick solid line).
}
\label{tagv_evt_dist}
\end{figure}
The probability density function (PDF) of signal is composed of
two parts:
one is for the candidates that are correctly reconstructed
combinations of 
daughter particles coming from a single neutral $B$ meson,
the other corresponds to combinations in which one of 
the final state particles is incorrectly
reconstructed (i.e. one of the daughter
particles originates from the other $B$ meson).
The former is parameterized by a two-dimensional function that 
is a product of a Crystal Ball line shape ~\cite{crystal_ball} 
in $\Delta E$ and a Gaussian form in $M_{\rm bc}$.
This parameterization accounts for the fact that 
$\Delta E$ and $M_{\rm bc}$ distributions are predominantly affected 
by the shower energy leakage in the ECL (in $\Delta E$) and 
the beam energy spread of the KEKB accelerator (in $M_{\rm bc}$).
On the other hand, the latter is described by a 
MC-determined two-dimensional smooth function.
In the signal box, the correct combination is estimated
to describe $87\pm2$ \% of the signal events.
%
%
The background is composed of four components: 
(1) $B^0\to J/\psi\, K^0_S$, 
(2) $B^0\to J/\psi\, K^0_L$, 
(3) $B\to J/\psi\, X$ other than $B^0\to J/\psi\, K^0$, 
(4) combinatorial background that consists of random combinations of particles
in $B\overline{B}$ decays and continuum events.
Using a large MC sample, the PDFs to describe (1), (2) and (3)
are determined and then parameterized as two-dimensional smooth functions
in $\Delta E\!-\!M_{\rm bc}$.
In the fit, we fix each yield of the three components, (1), (2) and (3),
to the values obtained from the MC sample.
The dominant $B\to J/\psi\, X$ contributions, 
excluding $J/\psi\,K^0_S$ and $J/\psi\,K^0_L$, 
come from two-body decays, 
with well-measured branching fractions
$(B\to J/\psi\, K^{(*)})$.
The combinatorial background shapes in $\Delta E$ and $M_{\rm bc}$
are described by a first-order polynomial and an ARGUS function~\cite{argus}, 
respectively.
The purity in the signal region
is estimated to be $87.9\pm8.0\%$.
The fractions of $J/\psi\, K^0_{S}$, $J/\psi\, K^0_{L}$ and
other $J/\psi\, X$ events are
$2.6\pm0.2\%$, $2.0\pm1.2\%$ and $3.2\pm0.2\%$, respectively,
while the combinatorial event fraction is $4.3\pm0.5\%$.
The $\Delta E$ and $M_{\rm bc}$ distributions after tagging and vertexing 
are shown in Fig.~\ref{tagv_evt_dist}.

We determine $\mathcal{S}_{J/\psi\, \pi^0}$ and 
$\mathcal{A}_{J/\psi\, \pi^0}$  by performing
an unbinned maximum-likelihood fit to the observed 
$\Delta t$ distribution:
\begin{equation}
\mathcal{L}(\mathcal{S}_{J/\psi\, \pi^0},\mathcal{A}_{J/\psi\, \pi^0})=
\prod^{N}_{i}\mathcal{P}(\mathcal{S}_{J/\psi\, \pi^0},\mathcal{A}_{J/\psi\, \pi^0};\Delta t_{i}),
\label{likelicpfiteq}
\end{equation}
where the product is over all events in the signal region.
The PDF
$\mathcal{P}$ is given by,
\begin{eqnarray}
\lefteqn{\mathcal{P}}~&=&(1-f_{\rm ol})\{
\int d(\Delta t') R(\Delta t_i-\Delta t')\mbox{\Large[}\nonumber\\
& &f_{\rm sig}\mathcal{P}_{\rm sig}(
\Delta t')
\nonumber\\
 & &+f_{J/\psi\, K_S}\mathcal{P}_{J/\psi\, K_S}(\Delta t')
+f_{J/\psi\, K_L}\mathcal{P}_{J/\psi\, K_L}(\Delta t')
\nonumber\\
 & &+f_{J/\psi\, X}\mathcal{P}_{J/\psi\, X}(\Delta t')
\nonumber
\mbox{\Large]}\nonumber\\
 & &+f_{\rm comb}\mathcal{P}_{\rm comb}(\Delta t_i)
\}\nonumber\\
   & &+f_{\rm ol}P_{\rm ol}(\Delta t_i),
\end{eqnarray}
where $f_{\rm sig}$, $f_{J/\psi\, K_S}$, $f_{J/\psi\, K_L}$, 
$f_{J/\psi\, X}$ and $f_{\rm comb}$
are the fractions of $B^0 \to J/\psi\, \pi^0$ signal, 
$B^0 \to J/\psi\, K^0_S$, $B^0 \to J/\psi\, K^0_L$, other $B\to J/\psi\, X$ 
background and combinatorial background, respectively.
All fractions are functions of $\Delta E$ and $M_{\rm bc}$
and are determined from the fit discussed above.
The PDF for the signal distribution, $\mathcal{P}_{\rm sig}$, is 
given by Eq.~\ref{signal_cp} and modified to account for
the effect of incorrect
flavor assignment;
the parameters $\tau_{B^0}$ and $\Delta m_d$ are fixed to 
PDG2006 values~\cite{pdg2006}.
The signal PDF is convolved with the proper-time 
interval resolution function $R(\Delta t)$~\cite{rsl}.
The $B^0 \to J/\psi\, K^0_S$
and $B^0 \to J/\psi\, K^0_L$ background
distributions are described by the same $\mathcal{P}_{\rm sig}$, 
respectively called $\mathcal{P}_{J/\psi\, K_{S}}$ and
$\mathcal{P}_{J/\psi\, K_{L}}$,
convolved with $R(\Delta t)$. The $CP$-asymmetry parameters
$\mathcal{S}_{J/\psi\, K^0_S}$, $\mathcal{A}_{J/\psi\, K^0_S}$,
$\mathcal{S}_{J/\psi\, K^0_L}$ and $\mathcal{A}_{J/\psi\, K^0_L}$
are fixed
to the recent Belle results~\cite{k0}.
The $B\to J/\psi\, X$ background excluding the $B^0\to J/\psi\, K^0_S$ and 
$B^0\to J/\psi\, K^0_L$ components
$(\mathcal{P}_{J/\psi\, X})$ is described 
with an effective lifetime as,
\begin{equation}
\mathcal{P}_{J/\psi\, X}(\Delta t)
  =\frac{e^{-|\Delta t|/\tau_{J/\psi\, X}}}
        {4\tau_{J/\psi\, X}}
		  \mbox{\huge\{}1-q\Delta \omega_l\mbox{\huge\}}.
\label{jxrmpdf}
\end{equation}
The effective lifetime $\tau_{J/\psi\, X}$ is
$1.10\pm0.10~(1.03\pm0.07)$ps
for the 3(4)-layer-silicon vertex detector sample,
which is determined by 
fitting a $B\to J/\psi\, X$ MC sample.
The combinatorial component ($\mathcal{P}_{\rm comb}$)
is described by a double Gaussian.
The relevant parameters are obtained using events in the sideband region,
5.20 GeV/$c^2 < M_{\rm bc} < 5.26$ GeV/$c^2$ and 
$|\Delta E| < 0.2$ GeV.
The fraction $f_{\rm ol}$ and PDF $\mathcal{P}_{\rm ol}$
describe the outlier component,
which is a small number
of events that have large $\Delta t$ values for both signal
and background.

\begin{figure}[htb]
\begin{center}
\includegraphics[width=7.5cm]{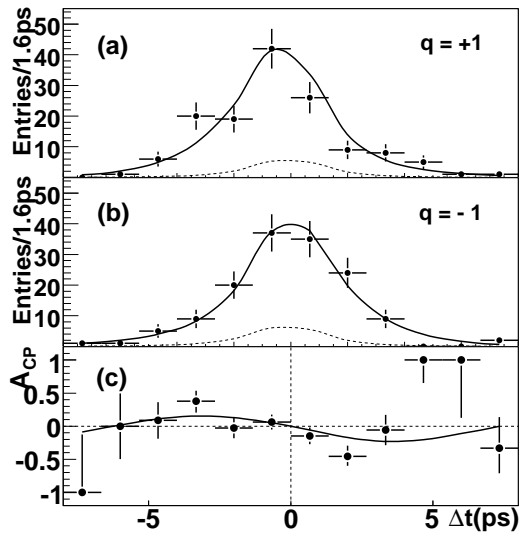}
\caption{$\Delta t$ distribution of $B^0\to J/\psi\, \pi^0$
candidate events for $q = +1$ (a) and $q = -1$ (b).
The dashed lines are the sum of backgrounds
while the solid lines are the sum of signal and backgrounds.
(c) is the raw asymmetry ($\mathcal{A}_{CP}$) distribution. The curve is 
the projection of the fit result.}
\label{cpfitFig}
\end{center}
\end{figure}

The unbinned maximum likelihood fit to the 290 events 
in the signal region results 
in the $CP$ violation parameters:
\begin{eqnarray}
\mathcal{S}_{J/\psi\, \pi^0} = -0.65\pm0.21(\mbox{stat})\pm0.05(\mbox{syst})\; {\rm and} \nonumber \\
\mathcal{A}_{J/\psi\, \pi^0} = +0.08\pm0.16(\mbox{stat})\pm0.05(\mbox{syst}),\nonumber
\end{eqnarray}
where the systematic uncertainties listed are described below.
The $\Delta t$ distributions and the time-dependent decay rate raw asymmetry $\mathcal{A}_{CP}$ are shown in 
Fig.~\ref{cpfitFig}, where $\mathcal{A}_{CP} = (N_+-N_-)/(N_++N_-)$
and $N_{+}~(N_{-})$ is the number of candidate events with $q=+1~(-1)$.
\begin{table}[htb]
\caption{Systematic uncertainties}
\label{systTable}
\begin{center}
\begin{tabular}{lcc}
\hline \hline
Parameter & $\Delta \mathcal{S}_{J/\psi\, \pi^0}$ 
& $\Delta \mathcal{A}_{J/\psi\, \pi^0}$  \\
\hline
Vertexing		                  & $\pm$0.050	& $\pm$0.034 \\
Wrong tag fraction	                  & $\pm$0.009	& $\pm$0.009 \\
Resolution function	                  & $\pm$0.008	& $\pm$0.007 \\
Fit bias		                  & $\pm$0.013  & $\pm$0.010 \\
Physics parameters		          & $\pm$0.004	& $\pm$0.001 \\
$B\to J/\psi\, X$ $CP$ asymmetry            & $\pm$0.004  & $\pm$0.001 \\
PDF Shape and fraction                    & $\pm$0.009	& $\pm$0.005 \\
Background $\Delta t$ shape	          & $\pm$0.006	& $\pm$0.001 \\
Tag side interference	                  & $\pm$0.001	& $\pm$0.038 \\
\hline
Total	                                  & $\pm$0.054	& $\pm$0.054 \\
\hline \hline
\end{tabular}
\end{center}
\end{table}

The systematic errors are listed in Table~\ref{systTable}.
The main contributions to the systematic error in 
$\mathcal{S}_{J/\psi\, \pi^0}$ are due to 
uncertainties in the vertex reconstruction 
and to a small fit bias.
The vertex reconstruction systematic error consists of uncertainties
in the interaction point profile, charged track selection based on the
track helix error,
helix parameter corrections, event selection based on $\Delta t$
and goodness of fit in the vertex reconstruction,
and the small SVD misalignment.
The systematic uncertainties due to the parameters $w_l$ and $\Delta w_l$
are estimated by varying the parameters by their one standard
deviation ($\sigma$) errors.
We vary each resolution function parameter by $\pm1\sigma$
and assign a systematic error as the quadratic sum of the
resulting deviations in $\mathcal{S}$ and $\mathcal{A}$.
The fit bias systematic error is evaluated from
an ensemble of MC samples as the difference between
the input and fitted values of $\mathcal{S}$ and $\mathcal{A}$.
The errors in the physics parameters $\tau_{B^0}$ and $\Delta m_d$ 
are taken into account.
To estimate the systematics from $B^0\to J/\psi\, K^0_S$
and $B^0\to J/\psi\, K^0_L$,
we vary their fractions and $CP$ asymmetry parameters,
$\mathcal{S}$ and $\mathcal{A}$ by $\pm1\sigma$.
We estimate the systematic uncertainty from the
$B\to J/\psi\, X$ backgrounds other than
$J/\psi\, K^0$ by scaling the systematic errors due to
$J/\psi\, K^0_S$ and $J/\psi\, K^0_L$
according to the amount of background contamination.
The $\Delta E$ and $M_{\rm bc}$ parameters and fractions of signal
and backgrounds are varied to estimate the systematic errors.
We vary the MC-determined parameters by $\pm2\sigma$ to
take account of possible imperfect modeling in MC.
We include the effect of tag side interference~\cite{taginter}.
The tag side interference is caused by the interference between the 
two amplitudes of $B$ decays into charmed mesons, 
i.e. caused by $V_{\rm cb}$ and $V_{\rm ub}$.
Therefore it is expressed by four parameters, 
$r_{\rm int}$ (size of interference between 
$V_{\rm cb}$ and $V_{\rm ub}$ amplitudes), 
$\phi_1$, $\phi_3$ and
$\delta$ (strong phase difference between 
$V_{\rm cb}$ and $V_{\rm ub}$ mediated amplitudes).
Since this interference results in a potential direct $CP$ violation, 
${\mathcal A}_{J/\psi \pi^0}$ is much more 
affected than ${\mathcal S}_{J/\psi \pi^0}$.
We sum each of the contributions in quadrature to obtain the total 
systematic error.

\begin{figure}[htb]
\begin{center}
\includegraphics[width=7.5cm]{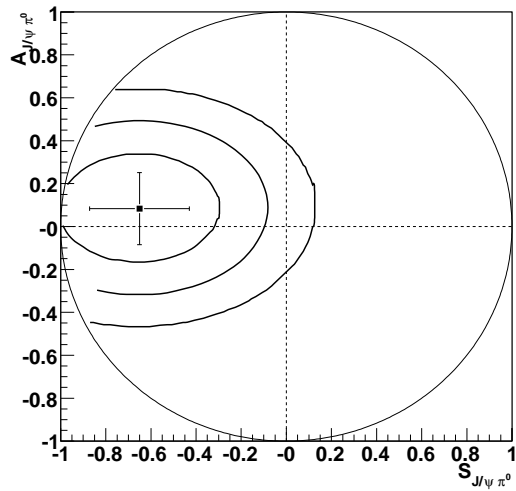}
\caption{
The confidence regions for 
$\mathcal{S}_{J/\psi\, \pi^0}$ and $\mathcal{A}_{J/\psi\, \pi^0}$.
The contours correspond to 1-C.L. = $3.17\times10^{-1}$ $(1\sigma)$,
$4.55\times 10^{-2}$ $(2\sigma)$, and $2.70\times 10^{-3}$ $(3\sigma)$.
The point with error bars corresponds to the measured
$\mathcal{S}_{J/\psi\, \pi^0}$ and $\mathcal{A}_{J/\psi\, \pi^0}$
values.
The circle is a boundary derived from Eq.(\ref{para_lambda}).
}
\label{conf_level}
\end{center}
\end{figure}

The confidence regions of our measurement in the 
$\mathcal{S}_{J/\psi\, \pi^0}$ and $\mathcal{A}_{J/\psi\, \pi^0}$ plane 
are shown in Fig.~\ref{conf_level}.
We evaluate the statistical significance of this $CP$ asymmetry measurement 
using a two-dimensional Feldman and Cousins method~\cite{feldmancousins}, 
taking both statistical and systematic uncertainties into account.
We found that our $\mathcal{S}_{J/\psi\, \pi^0}$ measurement has
a significance
greater than 2.4 $\sigma$ for any $\mathcal{A}_{J/\psi\, \pi^0}$ value.

In summary, we measure the $CP$ violation parameters 
in $B^0\to J/\psi\, \pi^0$ decays using $535\times 10^6 B\overline{B}$ pairs: 
$\mathcal{S}_{J/\psi\, \pi^0} 
= -0.65\pm0.21 (\rm{stat})\pm0.05 (\rm{syst})$ and
$\mathcal{A}_{J/\psi\, \pi^0} 
= +0.08\pm0.16 (\rm{stat})\pm0.05 (\rm{syst})$.
We measure mixing-induced $CP$ violation with 2.4 $\sigma$ significance.
This result supersedes our previous measurement~\cite{kataokaPRL} 
and exhibits significant improvement in precision compared to 
the latest BaBar measurement~\cite{babar}. 
It is consistent with the measured value of $\sin2\phi_1$
in $b\to c\overline{c}s$ decays~\cite{k0,k0_babar}, as expected
in the Standard Model.
%
%

We thank the KEKB group for the excellent operation of the
accelerator, the KEK cryogenics group for the efficient
operation of the solenoid, and the KEK computer group and
the National Institute of Informatics for valuable computing
and Super-SINET network support. We acknowledge support from
the Ministry of Education, Culture, Sports, Science, and
Technology of Japan and the Japan Society for the Promotion
of Science; the Australian Research Council and the
Australian Department of Education, Science and Training;
the National Science Foundation of China and the Knowledge
Innovation Program of the Chinese Academy of Sciences under
contract No.~10575109 and IHEP-U-503; the Department of
Science and Technology of India; 
the BK21 program of the Ministry of Education of Korea, 
the CHEP SRC program and Basic Research program 
(grant No.~R01-2005-000-10089-0) of the Korea Science and
Engineering Foundation, and the Pure Basic Research Group 
program of the Korea Research Foundation; 
the Polish State Committee for Scientific Research; 
the Ministry of Education and Science of the Russian
Federation and the Russian Federal Agency for Atomic Energy;
the Slovenian Research Agency;  the Swiss
National Science Foundation; the National Science Council
and the Ministry of Education of Taiwan; and the U.S.\
Department of Energy.

\end{document}